\PassOptionsToPackage{unicode}{hyperref}
\PassOptionsToPackage{hyphens}{url}
\PassOptionsToPackage{dvipsnames,svgnames,x11names}{xcolor}
\documentclass[
]{article}
\usepackage{amsmath,amssymb}
\usepackage{lmodern}
\usepackage{iftex}
\usepackage[T1]{fontenc}
\usepackage[utf8]{inputenc}
\usepackage{textcomp} 
\IfFileExists{upquote.sty}{\usepackage{upquote}}{}
\IfFileExists{microtype.sty}{
  \usepackage[]{microtype}
  \UseMicrotypeSet[protrusion]{basicmath} 
}{}
\makeatletter
\@ifundefined{KOMAClassName}{
  \IfFileExists{parskip.sty}{%
    \usepackage{parskip}
  }{
    \setlength{\parindent}{0pt}
    \setlength{\parskip}{6pt plus 2pt minus 1pt}}
}{
  \KOMAoptions{parskip=half}}
\makeatother
\usepackage{xcolor}
\providecommand{\tightlist}{%
  \setlength{\itemsep}{0pt}\setlength{\parskip}{0pt}}
\NewDocumentCommand\citeproctext{}{}
\NewDocumentCommand\citeproc{mm}{%
  \begingroup\def\citeproctext{#2}\cite{#1}\endgroup}
\makeatletter
 \let\@cite@ofmt\@firstofone
 \def\@biblabel#1{}
 \def\@cite#1#2{{#1\if@tempswa , #2\fi}}
\makeatother
\newlength{\cslhangindent}
\newlength{\csllabelwidth}
\newenvironment{CSLReferences}[2] 
 {\begin{list}{}{%
  \setlength{\itemindent}{0pt}
  \setlength{\leftmargin}{0pt}
  \setlength{\parsep}{0pt}
  \ifodd #1
   \setlength{\leftmargin}{\cslhangindent}
   \setlength{\itemindent}{-1\cslhangindent}
  \fi
  \setlength{\itemsep}{#2\baselineskip}}}
 {\end{list}}
\usepackage{calc}

\ifLuaTeX
\usepackage[bidi=basic]{babel}
\else
\usepackage[bidi=default]{babel}
\fi
\babelprovide[main,import]{american}

\def\languageshorthands#1{}
\ifLuaTeX
  \usepackage{selnolig}  
\fi
\IfFileExists{bookmark.sty}{\usepackage{bookmark}}{\usepackage{hyperref}}
\IfFileExists{xurl.sty}{\usepackage{xurl}}{} 
\hypersetup{
  pdftitle={SOAP: A Python Package for Calculating the Properties of
Galaxies and Halos Formed in Cosmological Simulations},
  pdfauthor={Robert McGibbon, John C. Helly, Joop Schaye, Matthieu
Schaller, Bert Vandenbroucke},
  pdflang={en-US},
  colorlinks=true,
  linkcolor={Maroon},
  filecolor={Maroon},
  citecolor={Blue},
  urlcolor={Blue},
  pdfcreator={LaTeX via pandoc}}

\title{SOAP: A Python Package for Calculating the Properties of Galaxies
and Halos Formed in Cosmological Simulations}

\definecolor{c53baa1}{RGB}{83,186,161}
\definecolor{c202826}{RGB}{32,40,38}

\usepackage[affil-it]{authblk}
\usepackage{orcidlink}
\author[1%
  ]{Robert McGibbon%
    \,\orcidlink{0000-0003-0651-0776}\,%
    }
\author[2%
  ]{John C. Helly%
    \,\orcidlink{0000-0002-0647-4755}\,%
    }
\author[1%
  ]{Joop Schaye%
    \,\orcidlink{0000-0002-0668-5560}\,%
    }
\author[1,3%
  ]{Matthieu Schaller%
    \,\orcidlink{0000-0002-2395-4902}\,%
    }
\author[1%
  ]{Bert Vandenbroucke%
    \,\orcidlink{0000-0001-7241-1704}\,%
    }

\affil[1]{Leiden Observatory, Leiden University, PO Box 9513, 2300 RA
Leiden, The Netherlands%
  }
\affil[2]{Institute for Computational Cosmology, Department of Physics,
University of Durham, South Road, Durham, DH1 3LE, UK%
  }
\affil[3]{Lorentz Institute for Theoretical Physics, Leiden University,
PO box 9506, 2300 RA Leiden, The Netherlands%
  }
\date{24 January 2025}

\begin{document}
\maketitle

\section{Summary}\label{summary}

Cosmological simulations model the evolution of dark matter and baryons
under the influence of gravitational and hydrodynamic forces. Beginning
at high redshift, they capture the hierarchical formation of structures,
where smaller structures form first and later merge into larger ones.
These simulations incorporate hydrodynamics to evolve the gas and
include a number of subgrid prescriptions for modelling important
physical processes, such as star formation. Once the simulation has
concluded, a halo-finding algorithm is used to identify bound structures
(subhalos) within the resulting particle distribution.

Here, we introduce \texttt{SOAP} (Spherical Overdensity \& Aperture
Processor), a Python package designed to compute halo and galaxy
properties from simulations that have been post-processed with a halo
finder. \texttt{SOAP} takes a subhalo catalogue as input and calculates
a wide array of properties for each object. Its output is compatible
with the swiftsimio package (\citeproc{ref-swiftsimio}{Borrow \&
Borrisov, 2020}), enabling seamless unit handling. \texttt{SOAP} has
already been used to generate halo catalogues for the FLAMINGO
simulation suite (\citeproc{ref-flamingoCalibration}{Kugel et al.,
2023}; \citeproc{ref-flamingo}{Schaye et al., 2023}), which includes the
largest cosmological hydrodynamic simulation to date. These catalogues
have been used in more than 20 publications to date.\footnote{For a
  complete list, see https://flamingo.strw.leidenuniv.nl/papers.html}

\section{Statement of Need}\label{statement-of-need}

Modern galaxy simulations are often analyzed by a large number of
researchers. However, due to the large volume of data generated, it is
often impractical for individual users to compute the specific
properties they require independently. \texttt{SOAP} addresses this
challenge by producing comprehensive catalogues containing a wide range
of properties that can be shared across the community.

Given the substantial volume of data, it is essential for the output to
be processed in parallel. \texttt{SOAP} achieves this using the mpi4py
library (\citeproc{ref-mpi4py}{Dalcín et al., 2005},
\citeproc{ref-mpi4py_2}{2008}, \citeproc{ref-mpi4py_3}{2011};
\citeproc{ref-mpi4py_4}{Dalcín \& Fang, 2021}). This enables
\texttt{SOAP} to scale efficiently across multiple compute nodes.
\texttt{SOAP} is also designed to handle subvolumes of the simulation
independently, allowing for large simulations to be processed
sequentially if required. This approach reduces the need for high-memory
resources. The ability to efficiently process subhalos in parallel is a
unique feature of \texttt{SOAP} when compared with other packages for
computing galaxy properties (e.g. \citeproc{ref-cosmic-profiles}{Dome,
2023}; \citeproc{ref-caesar}{Narayanan et al., 2023};
\citeproc{ref-tangos}{Pontzen \& Tremmel, 2018}).

A large number of halo finders are used by the community to identify
bound structures within simulation outputs. These employ a variety of
methods which can result in subhalo catalogues with significant
differences (\citeproc{ref-hbt-herons}{Forouhar Moreno et al., 2025}).
Therefore, it is important to be able to compare the results of various
halo finders to help quantify the uncertainty associated with structure
finding. However, along with the different structure identification
methods, the halo finder codes often vary in their implementation of
halo/galaxy property calculations and may even have different
definitions (e.g.~using inclusive/exclusive bound mass) for the same
property. This can lead to further differences in the resulting
catalogues, although in this case it is not due to the halo finding
method itself. \texttt{SOAP} can take input from multiple halo finders
and calculate properties consistently, thereby enabling actual
differences between structure-finding algorithms to be identified.
Currently \texttt{SOAP} supports HBT-HERONS
(\citeproc{ref-hbt-herons}{Forouhar Moreno et al., 2025};
\citeproc{ref-hbt}{Han et al., 2018}), SubFind
(\citeproc{ref-subfind}{Springel et al., 2001}), VELOCIraptor
(\citeproc{ref-velociraptor}{Elahi et al., 2019}), and ROCKSTAR
(\citeproc{ref-rockstar}{Behroozi et al., 2013}). Adding a new halo
finder requires a script to convert the subhalo catalogue into the
standard format used by SOAP; no other code changes are necessary.

The most common definition of a halo is based on spherical overdensities
(SO): regions of the universe which have a much larger density than the
average. The overdensity of a region is based on all the particles
within it, whether bound or unbound, and is therefore not always output
by halo finders. \texttt{SOAP} determines spherical overdensity radii by
constructing expanding spheres until the target density limit is
reached. It then calculates the properties of each halo using all the
particles within its SO radius. \texttt{SOAP} also calculates properties
for several other definitions of a halo: subhalo properties (using all
particles bound to a subhalo), fixed physical projected apertures (using
all bound particles within a projected radius), and two types of fixed
physical apertures (using all/bound particles within a sphere of the
same radius for all objects). These various types give users the freedom
to select the most appropriate definition for their scientific use case
e.g.~the type of observational data they are comparing with.

\section{Overview of Features}\label{overview-of-features}

\begin{itemize}
\tightlist
\item
  \texttt{SOAP} can currently calculate over 250 halo and galaxy
  properties. Users can easily add new properties to tailor the tool to
  their specific scientific needs. When combined with the four different
  halo definitions, this makes \texttt{SOAP} exceptionally versatile.
\item
  \texttt{SOAP} is compatible with both dark matter-only (DMO) and full
  hydrodynamic simulations. For DMO runs, any properties which are
  irrelevant (e.g.~gas mass) are automatically excluded, requiring no
  changes to the parameter file.
\item
  \texttt{SOAP} makes it easy to enable or disable specific halo
  definitions and properties using the \texttt{SOAP} parameter file.
  This is possible because all properties are lazily defined within the
  code and are only computed if required. Additionally, if certain
  objects require further analysis, \texttt{SOAP} can be run on a subset
  of subhalos.
\item
  Properties can be assigned filters so that they are only calculated
  for objects that meet certain criteria (e.g.~only calculate the halo
  concentration if a subhalo has a minimum number of bound particles of
  a particular type). This improves the runtime of \texttt{SOAP} and
  also reduces the data volume of the final output catalogues.
\item
  \texttt{SOAP} was originally written to run on Swift simulation
  snapshots (\citeproc{ref-swift}{Schaller et al., 2024}), utilizing
  their metadata for unit handling and spatial sorting to enable
  efficient loading of the data. However, it has also been used to
  create halo catalogues from the EAGLE simulation
  (\citeproc{ref-eagle}{Schaye et al., 2015}) snapshots (which use a
  modified GADGET format, \citeproc{ref-gadget}{Springel, 2005}).
  Supporting additional snapshot formats requires a conversion script to
  be written.
\item
  The output is saved as an HDF5 file which is spatially sorted,
  enabling quick loading of simulation subvolumes for analysis without
  requiring the entire dataset.
\item
  The catalogues can be read with the swiftsimio package
  (\citeproc{ref-swiftsimio}{Borrow \& Borrisov, 2020}), which provides
  unit conversion (including handling comoving versus physical
  coordinates) and a number of visualization tools. All datasets are
  output in units that are \emph{h}-free.
\item
  When provided with a parameter file, \texttt{SOAP} can automatically
  generate a corresponding PDF document with a detailed description of
  all the output properties. This ensures that the documentation of the
  generated catalogues (e.g., the property names, units, compression
  level, etc.) always reflects the specific setup of the current
  \texttt{SOAP} run.
\end{itemize}

\section{Acknowledgements}\label{acknowledgements}

We gratefully acknowledge contributions to the code from Joey
Braspenning, Jeger Broxterman, Evgenii Chaikin, Camila Correa, Victor
Forouhar Moreno, and Roi Kugel. \texttt{SOAP} relies heavily on the
following packages: mpi4py (\citeproc{ref-mpi4py}{Dalcín et al., 2005},
\citeproc{ref-mpi4py_2}{2008}, \citeproc{ref-mpi4py_3}{2011};
\citeproc{ref-mpi4py_4}{Dalcín \& Fang, 2021}), NumPy
(\citeproc{ref-numpy}{Harris et al., 2020}), h5py
(\citeproc{ref-h5py}{Collette, 2013}), and unyt
(\citeproc{ref-unyt}{Goldbaum et al., 2018}).

\section*{References}\label{references}
\addcontentsline{toc}{section}{References}

\phantomsection\label{refs}

\end{document}